\documentclass{article}

\usepackage{arxiv}
\usepackage[utf8]{inputenc} 
\usepackage[T1]{fontenc}    
\usepackage{hyperref}       
\usepackage{url}            
\usepackage{booktabs}       
\usepackage{amsfonts}       
\usepackage{nicefrac}       
\usepackage{microtype}      
\usepackage{lipsum}		
\usepackage{graphicx}
\usepackage{doi}

\usepackage{caption}
\usepackage{subcaption}
\usepackage{algpseudocode}
\usepackage{algorithm}
\usepackage{listings}
\usepackage{amsmath}
\usepackage{xcolor}
\usepackage{listings}

\lstdefinestyle{customc}{
    language=C,                      
    basicstyle=\small\ttfamily,       
    keywordstyle=\color{blue},        
    commentstyle=\color{gray},        
    stringstyle=\color{red},          
    numbers=left,                     
    numberstyle=\tiny\color{gray},    
    backgroundcolor=\color{lightgray!20}, 
    frame=single,                     
    tabsize=4,                        
    breaklines=true                    
}

\newcommand{\frameworkName}{VIC } 

\title{VIC: Evasive Video Game Cheating via Virtual Machine Introspection}

\author{ \href{https://orcid.org/0000-0002-3110-019X}{\includegraphics[scale=0.06]{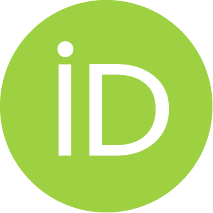}\hspace{1mm}Panicos Karkallis} \\
	Royal Holloway University of London\\
	Egham, United Kingdom \\
	\texttt{panicos.karkallis.2018@rhul.ac.uk} \\
	\And
	\href{https://orcid.org/0000-0003-4392-9023}{\includegraphics[scale=0.06]{orcid.pdf}\hspace{1mm}Jorge Blasco Alís} \\
        Computer Systems Department \\
	Universidad Politécnica de Madrid\\
	Madrid, Spain \\
	\texttt{jorge.blasco.alis@upm.es} \\
}

\renewcommand{\shorttitle}{\textit{VIC} Virtual Machine Introspection Cheat }

\hypersetup{
pdftitle={VIC - Evasive Video Game Cheating via Virtual Machine Introspection},
pdfsubject={q-bio.NC, q-bio.QM},
pdfauthor={Panicos Karkallis, Jorge Blasco Alis},
pdfkeywords={Online video game hacking, video game security, virtual machine introspection},
}

\begin{document}
\maketitle

\begin{abstract}
	Video game cheats modify a video game behaviour to give unfair advantages to some players while bypassing the methods game developers use to detect them. This destroys the experience of online gaming and can result in financial loses for game developers. In this work, we present a new type of game cheat, Virtual machine Introspection Cheats (VIC), that takes advantage of virtual machines to stealthy execute game cheats. \frameworkName employees a hypervisor with introspection enabled to lower down the bar of cheating against legacy and modern anti-cheat systems. We demonstrate the feasibility and stealthiness of \frameworkName against three popular games (Fortnite, BlackSquad and Team Fortress 2) that together include five different anti-cheats. In particular, we use \frameworkName to implement a cheat radar, a wall-hack cheat and a trigger-bot. To support our claim that this type of cheats can be effectively used, we present the performance impact that \frameworkName has on gameplay by monitoring the frames per second (fps) while the cheats are activated. Our experimentation also shows how these cheats are currently undetected by the most popular anti-cheat systems, enabling a new paradigm that can take advantage of cloud infrastructure to offer cheating-as-a-service.
\end{abstract}

\keywords{Online video game hacking, video game security, virtual machine introspection}
\textbf{\textit{ACM-Class: }}D.4.6

\section{Introduction}

Video game cheats are designed to give cheaters an unfair advantage over legitimate players in online games. 

To construct these, cheat developers employ a number of strategies. They manipulate the game's memory by reading and writing state variables which under normal circumstances are unavailable to them. They can also employ trampoline functions to redirect the game process execution and execute malicious code inside the game's memory \cite{FengStealthMeasurements}. Other techniques include monitoring network packets for information, and more recently computer vision was used to detect enemy avatars by reading the screen using a cheat known as PixelBot \cite{pixelbot}. 

Despite the use of sophisticated anti-cheat systems, cheat developers are constantly developing bypasses and stealthy cheating methods, resulting in a cat and mouse game where every new anti-cheat method is quickly bypassed by a new cheat. This has resulted in several layers of protection being added to modern video games to protect them from cheaters. Currently, the plethora of anti-cheat systems have extended their reach to kernel space as a response to defend against cheats which employee kernel drivers to access the game's process memory, as demonstrated by Joel Noguera \cite{unveilingUndergroundWorldAntiCheats}. This move might seem as the next natural step to take, although it comes with certain drawbacks: it can affect the operating system's performance, stability and security if not implemented correctly. One example of such failure is the Digital Rights Management System (DRM) that Sony BMG included within their CDs from 2005. This DRM was installing a rootkit on the user's OS in order to modify the operating system's behavior to stop CD copying  \cite{halderman2006SonyRootkit}.

This paper demonstrates a new vector of attack to enable video game cheats: Virtual machine Introspection Cheats (VIC). We show how direct memory and interrupt access to a guest OS can allow cheaters to bypass user-level and kernel-level anti-cheats. This is of particular relevance now that we have reached a point where broadband speeds and virtualization technologies are enabling a new way of accessing graphically advanced video games via streaming (as proven by services such as Xbox Cloud Gaming from Microsoft or GForce Now from Nvidia). We demonstrate our method with an implementation on three popular online games with millions of players worldwide: Team Fortress 2, BlackSquad and Fortnite. To the best of our knowledge, this is the first work that demonstrates the use of virtual machine introspection to execute video game cheats. In particular our contributions are:
\begin{itemize}
    \item We introduce a new method of cheating using a hypervisor with introspection enabled. We propose two approaches to achieve effective cheats by using a polling mechanism on the guest memory and by using page guard exceptions to assess the efficiency and applicability of this design paradigm.
    \item We use our method to implement three different types of cheats for three different games. With this, we demonstrate the effectiveness of our method to bypass both a user-level and a kernel level anti-cheats in current commercial video games.
    \item We discuss how our work could be used to help anti-cheat analysts and discuss the drawbacks involved when trying to defend against this type of cheats and mitigation's proposed in literature. 
    (\S\ref{sec:discussion}).
   
\end{itemize}

The rest of the paper is organized as follows.
We present the methodology used by \frameworkName in \S\ref{sec:vmiCheat}. 
We evaluate \frameworkName against three popular games and present our results in  \S\ref{sec:evaluation}. We discuss the implications and the limitations of our attack in \S\ref{sec:discussion} and, finally, summarize our conclusions in \S\ref{sec:conclusions}.

\section{Background and Related Work}
\label{sec:Background}

For the purpose of this paper, video game cheating can be defined as the act of gaining an advantage over other players by modifying or accessing information inside the game's process which were unintended by the game developers. Network packet modification and server or client side statistical analysis are both outside of the scope of our paper and therefore not included within our definition. However, these kinds of cheats have also been explored and used in the past \cite{BURSZTEIN2011OPENC,BURSZTEIN2016IAM,Lehtonen2020}. Cheats that conform with our definition are typically categorised based on two different criteria: how they get injected into the game, inside the game's process or outside, and what kind of advantage they provide to the cheater.

Modifying the video game behaviour normally requires changes on how the process running the video game is executing (changes in its memory space). \textbf{Internal cheats} are directly injected inside the game's process. When a cheat or any DLL library gets mapped into a process memory, it has access to game's functions and memory. Function hooks and trampoline functions are also used to introduce new behaviours as described by Feng et.al \cite{FengStealthMeasurements}. Karkallis et.al highlighted in their underground cheat forum analysis the importance and popularity of generic injectors and how they are one of the most widely used methods to inject cheats into video games \cite{DetectingGameInjectors}. \textbf{External cheats} exist outside the game's process. They can either exist in a process created by the cheat or get injected in elevated processes to take advantage of the additional privileges granted by the elevated process. This type of cheats normally have arbitrary access (read and write) to all the process running the game unlike internal cheats which are restricted by where they are injected.

When successfully executed, cheats provide unfair advantage to the players by several means. \textbf{Cheat Radars} map the position of enemy players in real time in a 2D virtual map of the video game world. This information helps the cheater to anticipate attacks by players in close proximity and creates an advantage over other players. The cheat takes advantage of the location information shared with the game client by the game server. In some cases the radar is drawn by using the cheater’s player position as point of reference. \textbf{Wall-hacks} highlight elements on screen so they can be easily identified by the cheater. Usually the highlighted parts are enemy avatars as means to understand their location before they get closer to the cheater's player. It involves a world-to-screen translation of the player's location to a 2D screen dimension.

 \textbf{Trigger-bots} reduce the cheaters reaction time in FPS games by automatically firing at an enemy when they are on target. They are considered more advanced compared to the two previous cheats as they do not only read the game state. They also process aiming information from the cheater’s avatar cross-hair and if an enemy avatar is on target it then commands the game client to start firing while the enemy avatar is in range.
\textbf{Aim-bots} increase shooting accuracy by automatically aiming at enemy avatars. The cheater's camera rotation gets adjusted accordingly to keep the cross-hair always on target of the enemy avatar. This cheat can be used in conjunction with a \textbf{No-recoil} cheat, which removes the recoil action when firing a weapon in shooting games thus increasing the players shooting accuracy. Depending on where the game state is kept cheats like \textbf{Speed-hacks} (increases the speed of the cheater's avatar) or \textbf{God-modes} (provides the player with invincibility) can also be enabled.

\textbf{Anti-cheats} are programs developed by game studios or third party companies to protect the game's processes from malicious activity/cheats.  At the time of writing this paper there exist three different types of client-side anti-cheats: User-mode, kernel-mode and anti-tamper. User-mode and kernel-mode anti-cheats monitor the game process and try to detect malicious interactions a cheat would try to do. 
Their main difference, as suggested by their name, is that kernel-level anti-cheats also employ a driver to monitor kernel space. User-level anti-cheats exist only in the user-space with elevated privileges and monitor the game's process through the available methods and resources available. Similar to user-level anti-cheats, anti-tamper anti-cheats can also monitor the game process for malicious activity but they take a preemptive strategy towards cheating. Their main goal is to increase the difficulty of reverse engineering the game binary. They obfuscate the game binary and encrypt parts of it, even while the game is running, so it cannot be analyzed by disassemblers \cite{androidObfuscation,balakrishnan2005code,you2010malware}. A more detailed description on anti-cheat methods and system comparison can be found in \cite{bohnerthanti}.

To help them create new cheats, cheat developers created \textbf{tools} to assist then in the task of narrowing down the parts of the game memory more useful to them. Two popular tools are CheatEngine\cite{cheatEngine} and ReClass\cite{reClass}. They both have many abilities including memory scanning and memory editing. CheatEngine, a widely-used tool for game hacking, in more recent years introduced a Type-1 hypervisor called Dark Byte's Virtual Machine (DBVM). The tool implements DBVM as a means of avoiding kernel-level checks by elevating memory accesses of the CheatEngine process using the hypervisor. There exist two ways to load DBVM onto a system. The first requires the cheater to load a driver onto the system which can be detectable by kernel anti-cheat systems. The second is to load DBVM from a universal serial bus (USB) stick at boot time alongside the OS.

\begin{figure} [H]
   
    \includegraphics[scale=0.5]{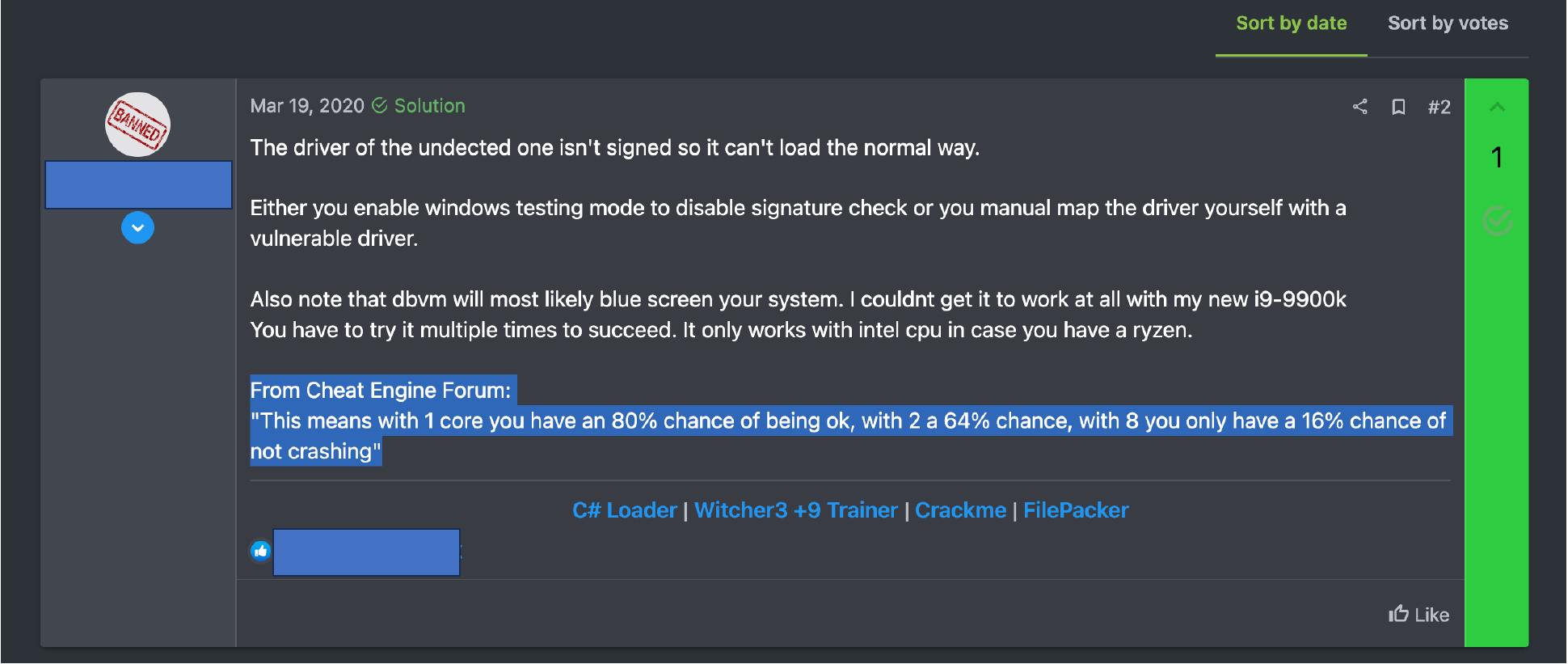}
    \caption{GuidedHacking forum post discussing DBVM crashes. \cite{dbvmcrash3}}
    \label{fig:dbvm_crash}
\end{figure}

DBVM, being a Type-1 hypervisor, comes with certain limitations. One significant issue is that the Cheat Engine process or any cheat related process e.g. overlays must operate within the same OS as both the game and its anti-cheat system. This arrangement compromises the stealthiness of the approach, especially since many anti-cheat systems maintain a blacklist of prohibited programs and prevent the game from running if these programs are active. Another challenge posed by the nature of a Type-1 hypervisor is that DBVM is more complex to test and modify. It also tends to be less stable, as reported by users on various online forums \cite{dbvmcrash1,dbvmcrash2,dbvmcrash3}. To illustrate this, Figure \ref{fig:dbvm_crash} shows a discussion from a Guided Hacking forum user commenting on the instability problems experienced with DBVM.

Type-2 hypervisors, such as Quick Emulator (QEMU), run as programs inside a host OS, for example Ubuntu Linux. In the realm of game cheating, this allows for the utilisation of the host OS to operate cheat-related processes, effectively using the hypervisor as a shield between the anti-cheat systems and the VIC. This setup eliminates the need for custom boot loaders that are necessary to run guest VMs in solutions like DBVM. Moreover, Type-2 hypervisors are generally more robust compared to DBVM. Another advantage of Type-2 hypervisors is their ability to emulate input devices, which can simulate player input in games. In practical applications, there have been discussions about the use of Type-2 hypervisors in online forums. There are relatively simple cheats observed in the wild that use a Type-2 hypervisor, mainly for reading and writing to game memory and debugging games \cite{guidedHackingVMIntro,uncHyperV,pareidoliatriggerbotHypervisor}. This creates the need for investigating further how more advanced VIC cheats can be implemented and what performance they might have.

 Various studies in the literature explore how VMs and virtualisation technologies are used to develop cheats, highlighting the challenges in detecting such cheats when a hypervisor is involved \cite{FengStealthMeasurements,lehtonen2020comparativeAntiCheats,spider}. However, none of these studies demonstrate the actual implementation of VIC, or show how they can be effectively used to gain an advantage over other players in online games.

This work aims to expand upon the understanding and the limitations of already implemented solutions e.g. DBVM and the gaps in the literature. The chapter evaluates the effectiveness of VIC using a Type-2 hypervisor with introspection capabilities across various popular games. It outlines a method for developing VIC cheats that minimally affect game performance. This approach takes advantage of the isolation provided by the hypervisor and the ability to emulate input, thereby ensuring that the cheats operate subtly. Moreover, the chapter also presents strategies for detecting these types of cheats, offering a comprehensive view of both the implementation and mitigation of VIC in the gaming context.

Our proposed attack in \S\ref{sec:vmiCheat} can be considered an external cheat as the cheat not only runs outside the game's process but outside the game's operating system. This division means anti-cheats cannot upload the cheat module injected to the game process to their servers for post analysis, something the majority of the anti-cheats use as means of understanding how new cheats work. It also offers an abstraction to cheat developers that facilitates their development as the same cheat can run on different anti-cheats with minor modifications apart from the update of memory object offsets. 


\section{Virtual Machine Introspection Cheats (VIC)}
\label{sec:vmiCheat}

\frameworkName takes advantage of modern virtualization extensions that allow CPU and GPU virtualization to be used as an enabler to run video games at optimal speed, even within virtual machines. We propose to run the video games inside a virtualized environment and use the hypervisor, and virtual machine introspection (VMI) in particular, to modify the guest machine to achieve the same end-result as a cheat that would be executed within the machine \cite{garfinkel2003virtual}.

\subsection{Threat model}
\label{sec:threatModel}

\begin{figure}[H]
    \centering
    \includegraphics[scale=0.40]{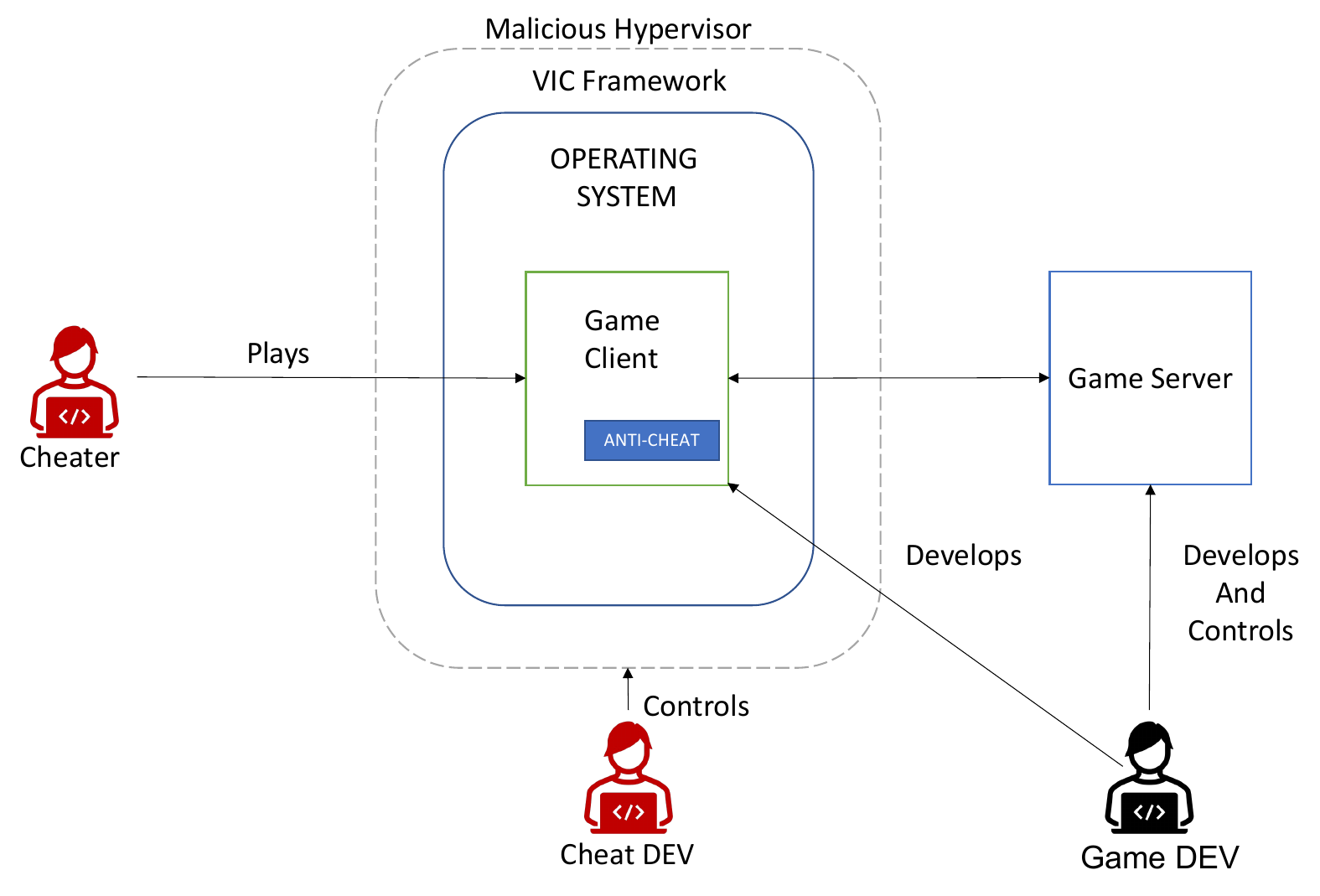}
    \caption{\frameworkName framework threat model.}
    \label{fig:threat-model}
\end{figure}

Our threat model involves three main actors, the cheater, the cheat developer and the game developer (Figure \ref{fig:threat-model}). The cheater is a game user who wants to play a particular game with a selection of cheats enabled. They can own the system where the cheat is running or access it in a similar fashion they would access any other cloud gaming service. The cheat developer is in charge of developing the cheats and enabling them on the game. They have full control over the system that will run the game with the cheats. This could be because the cheater has granted them full access or because they own the system and offer access to it as a service. In particular, we assume the cheat developer can modify the operating system (including running custom kernel versions) and is able to run a malicious hypervisor with introspection enabled which can host a guest VM on that system. In our evaluation, performed in \S\ref{sec:evaluation}, we assume that both the cheater and the cheat developer are the same person. This is, they own the system where the game is running, have full access to it and also want to play the game with cheats enabled. This is not necessarily always the case as we explain in \S\ref{sec:discussion} where the cheat developer can offer, via cloud infrastructure, direct online access to a cheat-enabled cloud gaming system. 

The game developer creates the video game that will be abused by the cheat developer. The game is meant to be played online but has a client that needs to be executed for every player. While most information is stored in the servers, a certain amount of information is needed at the client side for speed and responsiveness. This is consistent with how most current multiplayer online video games are played on personal computers. The game includes an anti-cheat that can have access to both, the user and kernel space of the system where the game is being executed. The video game is security agnostic, thus only relevant information are shared between the client (user's PC) and the server. The server also runs statistical analysis tools to detect cheating behaviours.

\subsection{Overview}
Cheats communicate with the game process through memory reads (to obtain information that would be useful to the cheater), function hooks (to modify specific behaviours or act on specific game events) and memory writes (to modify states such as firing).

Our proposed methodology translates the above actions and implements the same effects by placing the cheat outside the VM where the game is hosted. Reads/writes to memory take place through VMI and page guards are introduced to monitor specific memory addresses and specific code execution. In contrast to traditional cheats, we have the ability to employ IO device simulation to create mouse and keyboard events which act as a stealthier way to modify game state. 
Mapping the above actions is very important, as all the already implemented cheats on any game can run using our framework with minor modifications on the actions described above. In addition, the cheat will not need to use any anti-cheat bypass or custom driver, something a kernel cheat would need in order to exploit the game thus decreasing the barrier for cheating even in games which employ a number of anti-cheats like Fortnite.

\begin{figure}[H]
    \centering
    \includegraphics[scale=0.8]{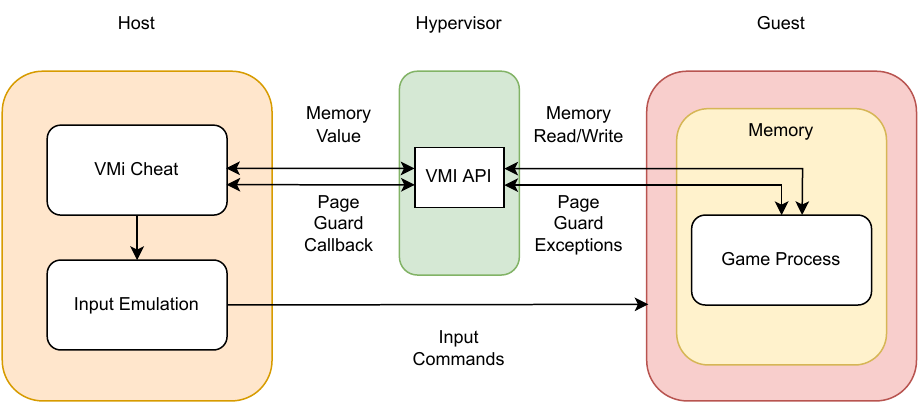}
    \caption{Overview of VIC attack pipeline.}
    \label{fig:methodology}
\end{figure}

In this work we use QEMU emulator with introspection enabled via LibVMI \cite{payne2012simplifying}, to demonstrate how the use of introspection can lower the bar for cheat developers and provide them an advantage over other players. Figure \ref{fig:methodology} demonstrates the flow of commands starting from the host on left hand side. The host machine runs the hypervisor and the VMI cheat. The cheats use LibVMI to communicate with KVM's introspection library. KVMI provides the cheat program direct read and write access to the guest VM's memory and access to different hardware interrupts e.g. page guard exceptions. The guest runs a vanilla operating system installation without any modifications, including the game installations. 

\subsection{Building Blocks}
\label{sec:building_blocks}

The use of LibVMI for the development of the \frameworkName is important as it opens the vector of attack not only to KVM but also the Xen hypervisor which provides even more stealthier options to monitor guests by employing altp2m (multiple EPTs) \cite{lengyel2016stealthy}. Many anti-cheats e.g. Easy Anti-Cheat and BattleEye do not let the game start if they detect kernel debugging is enabled. The support for volatility debug kernel symbols by LibVMI makes debugging on the guests kernel obsolete. We now describe how each of the different tasks required to build a video game cheat can be achieved via VMI:
\begin{itemize}
    \item \textbf{Reading} memory to the guest physical memory is achieved using the Second Level Address Table (SLAT) implemented using Extended Page Table (EPT) or Nested Page Tables (NPG). LibVMI walks the page tables of the guest VM by using the page directory location of the relevant process and reads the relevant memory pages and their address values from the host physical address.
    
    \item \textbf{Writing} to memory happens in a similar manner to reading. LibVMI walks the page tables of the guest VM by using the page directory location of the relevant process and updates the host physical address pointed by SLAT.
    
    \item \textbf{Monitoring} access to a value in memory is achieved via Page guard exceptions. To achieve a page guard exception or page fault, LibVMI modifies the guest page frame number (gfn) bits of the memory page which contains the address we want to receive exceptions for. Every time there is an exception on any address contained inside the memory page frame an associated callback method will get called. The callback gets called only when there is a relevant address modification. The performance of this method is bound to the number of events the page frame generates. This technique becomes particularly useful when the cheat needs to act on an in game event.
        
    \item To achieve \textbf{input emulation} we use the QEMU Machine Protocol (QMP), a JSON based protocol for communicating to QEMU machines using sockets. It enables the host to trigger a range of IO events including mouse clicks and key presses. QMP IO events are indistinguishable from IO events, triggered manually via a keyboard or a mouse, at the guest level.

    \item \textbf{Overlays} are transparent windows with the same dimensions as the game window which gets overlapped on top of the game window. They are mainly used to highlight different elements of the video game screen with to increase the player's advantage (e.g. position of an enemy player even if they are behind a wall). This normally gets implemented in different ways e.g. internal cheats hook either D3DX or OpenGL functions responsible for drawing game frames to the screen \cite{cano_2016}. In \frameworkName the overlay window runs on the host machine (outside the VM running the game). This makes it impossible to detect by current anti-cheat systems.
\end{itemize}

Using these blocks, we are capable of building a range of cheats that bypass current anti-cheat systems. In particular, we build \frameworkName to include three different cheats: a cheat-radar, a wall-hack and a trigger-bot that are evaluated in \S\ref{sec:evaluation}. 

\subsection{Selected Cheats}
\label{subsec:cheatcategories}

This section describes the implementation of the three cheats selected for \frameworkName. A more detailed description of how these cheats affect the game can be found on \S\ref{sec:Background}. Some of the cheats developed for this work have been partially based on sources by cheat developers that are posted online as means to gain reputation in online cheating communities, as mentioned in \cite{DetectingGameInjectors}. We leverage these already public cheats also to prove that our framework can be used to successfully translate traditional cheats into VMI-based cheats.

\begin{figure}[H]
     \centering
     \begin{subfigure}{0.45\textwidth}
         \includegraphics[width=\textwidth]{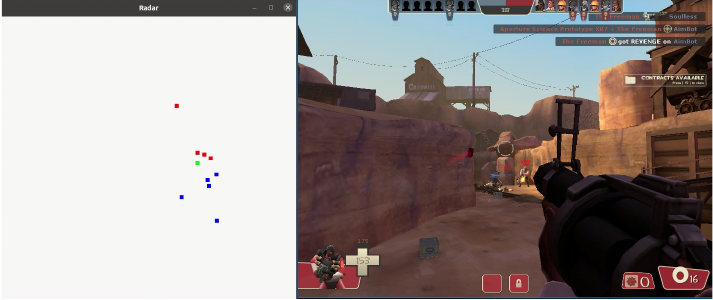}
         \caption{Team Fortress 2}
         \label{fig:tf2-cheat-radar}
     \end{subfigure}
     \hfill
     \begin{subfigure}{0.45\textwidth}
         \includegraphics[width=\textwidth]{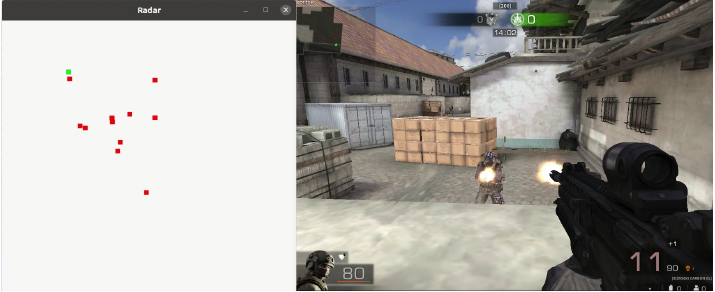}
         \caption{BlackSquad}
         \label{fig:bs-cheat-radar}
     \end{subfigure}
     \hfill
     \begin{subfigure}{0.45\textwidth}
         \centering
         \includegraphics[width=\textwidth]{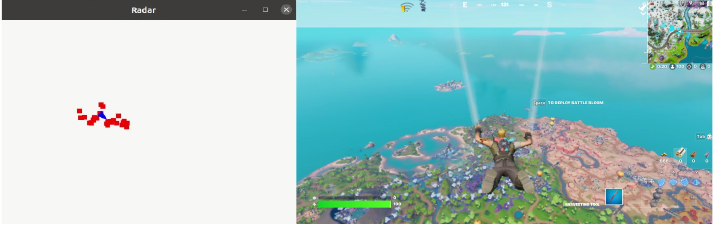}
         \caption{Fortnite}
         \label{fig:frt-chear-radar}
     \end{subfigure}
     
        \caption{Cheat radars for each of the games used during the evaluation using \frameworkName. The left-hand side shows the cheat radars for each game. The right-hand side is the actual game frame illustrated by the cheat radars.}
        \label{fig:cheat-radar}
\end{figure}

Similarly to "traditional" external cheats the cheat developer first needs to identify where the structures and information the cheat needs to function are stored in memory. This first discovery phase of the necessary offsets is done by reverse engineering the game binary using disassemblers (IDA \cite{ida}, Ghidra \cite{ghidra}) or by using memory scanning tools like Cheat Engine \cite{cheatEngine}. An easier way to find offsets is to use special programs called \emph{Dumpers}. Dumpers are usually developed for a specific game or game engine and try to automate the process of finding the relevant offsets. Another way to find offsets is to visit game hacking forums like \textit{MPGH} and \textit{UnknownCheats} in hope some other user has updated the game's forum with the latest offsets.  

\frameworkName uses two threads for any cheat that requires an overlay. The first thread (main thread) reads information from the game, while the second thread is responsible for rendering the required information on the screen (which depends on the cheat), \textit{Algorithm \ref{alg:uiThread}}.  
The first thread comes in two versions, \textit{Algorithm \ref{alg:mainThread} \& \ref{alg:mainThreadV2}}. Version \ref{alg:mainThread} is used when we need to read from memory in certain intervals. We do this by reading the corresponding memory position and sleeping until the next read. This can sometimes be inefficient as it requires the process to sleep, wake up, read the address again and compare the value to understand if there was a change. This results in unnecessary reads. 

Because of this, we introduce version \ref{alg:mainThreadV2}, a more efficient version that takes advantage of page guard exceptions. In this version we record the memory addresses that we want to monitor via page guard exceptions. We want to monitor these memory addresses because they may hold relevant information about the game state such as the position of an enemy avatar with respect to player. For each of those memory addresses then we add callbacks that execute the particular action we want the player to execute automatically when that memory address is accessed (the previously identified condition is met). This allows us to do a more efficient implementation of some cheats. 

However, this method has to be used carefully. If many memory addresses are monitored, the amount of page guard exceptions will result in a high number of events which can bring the guest VM to a halt. Because of this, our cheat implementation uses version \ref{alg:mainThread} and we have also produced a cheat implementation for the trigger-bot for Team Fortress 2 using version \ref{alg:mainThreadV2}. It is important to highlight that the \frameworkName runs only on the host machine as seen on Figure \ref{fig:methodology}.

\begin{algorithm}
\caption{\frameworkName Main Thread version a. This thread is in charge of obtaining the relevant information from game memory and passing it to the user interface thread.}\label{alg:mainThread}
\begin{algorithmic}
\State $gameValues \gets Array[PlayerObject, PlayerObject,...]$
\State $uiThread \gets \Call{$Thread$}{$gameValues$}$ \algorithmiccomment{Passes reference of array to Thread}
\State $uiThread.start()$ \algorithmiccomment{Starts User Interface Thread}
\While{$gameIsRunning$} 
  \State \Call{$UpdateGameValues$}{$gameValues$}
  \State \Call{$Sleep$}{$TimeInMilliseconds$}

\EndWhile
\end{algorithmic}
\end{algorithm}

\begin{algorithm}
\caption{\frameworkName Main Thread version b. This thread is in charge of monitoring a small number of addresses and execute callbacks on address change.}\label{alg:mainThreadV2}

\begin{algorithmic}
\Procedure{CallbackA}{$newAddressValue$}
  \State \Call{jump}{ }
\EndProcedure

\Procedure{CallbackB}{$newAddressValue$}
  \State \Call{startShooting}{ }
\EndProcedure
\Procedure{\textbf{main}}{}
\State $addressList \gets Array[int, int]$

 \State \Call{$registerException$}{$addressList[0]$, CallbackA}
 \State \Call{$registerException$}{$addressList[1]$, CallbackB}

 \State \Call{$ListenForExceptionEvents$}{ }
\EndProcedure
\end{algorithmic}
\end{algorithm}

\begin{algorithm}
\caption{User Interface Thread. This thread will create the overlay required for a particular cheat based on the game state information received from the main thread.}\label{alg:uiThread}
\begin{algorithmic}
\State $gameValues \gets mainThreadGameValues$
\While{$gameIsRunning$} 
  \State \Call{$Draw$}{$gameValues$}
    \State \Call{$Sleep$}{$TimeInMilliseconds$}

\EndWhile
\end{algorithmic}
\end{algorithm}

\subsubsection{Cheat radar.\label{subsec:cheat-radar}}Cheat radar requires the knowledge of the player's coordinates in the world space where all the game objects exist. These are obtained by reading memory at particular offsets found during the initial discovery phase. The x and y coordinates representing the player's position on the actual game map get scaled down to fit the dimensions of the cheat radar 2D map. In order to scale them down, the cheat developer first needs to identify the maximum value they can take. This can be accomplished by identifying the map dimensions for each game and select a maximum value accordingly. 

\subsubsection{Wall-hack. \label{subsec:wall-hack}}
\begin{figure}[H]
    \centering
    \includegraphics[scale=0.25]{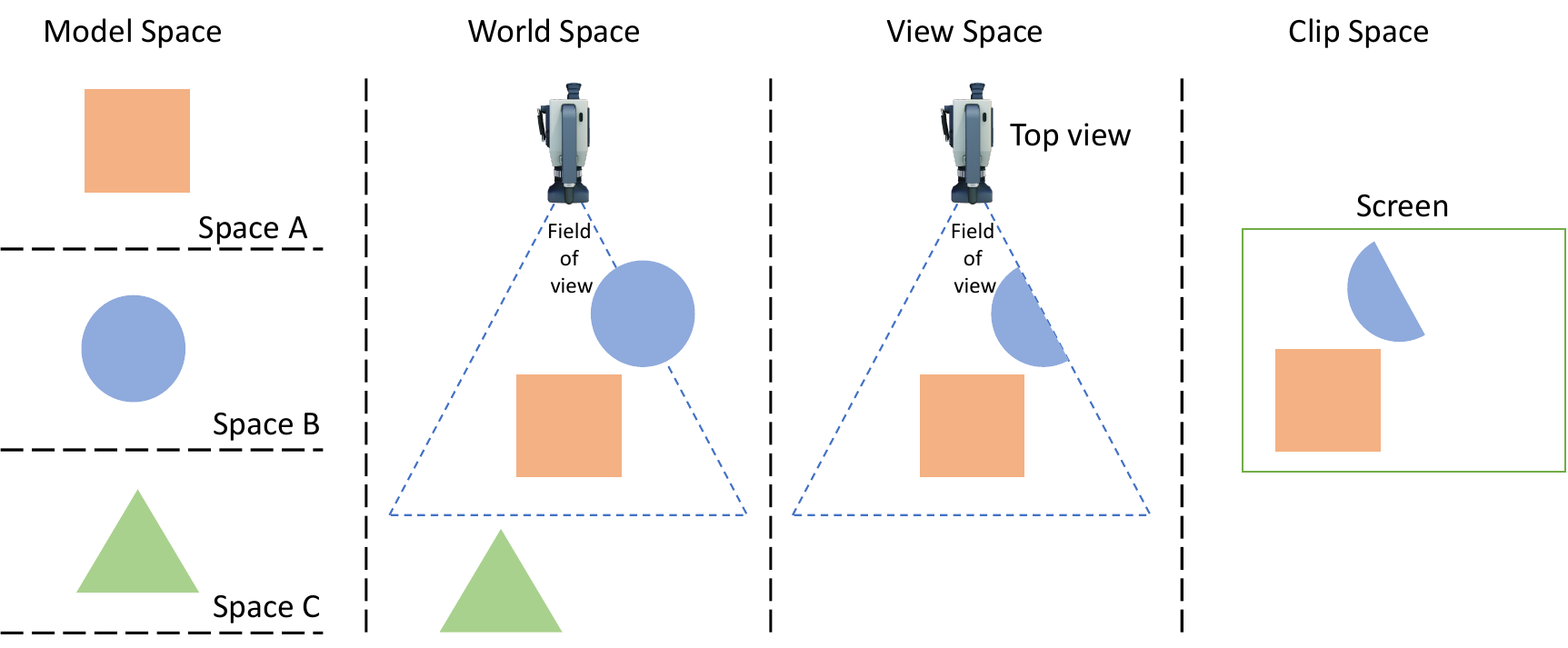}
    \caption{Illustration of model space to screen dimension transformation used for the wall-hack cheat.}
    \label{fig:world-to-screen}
\end{figure}

\begin{figure}
     \centering
     \begin{subfigure}{0.45\textwidth}
         \includegraphics[width=\textwidth]{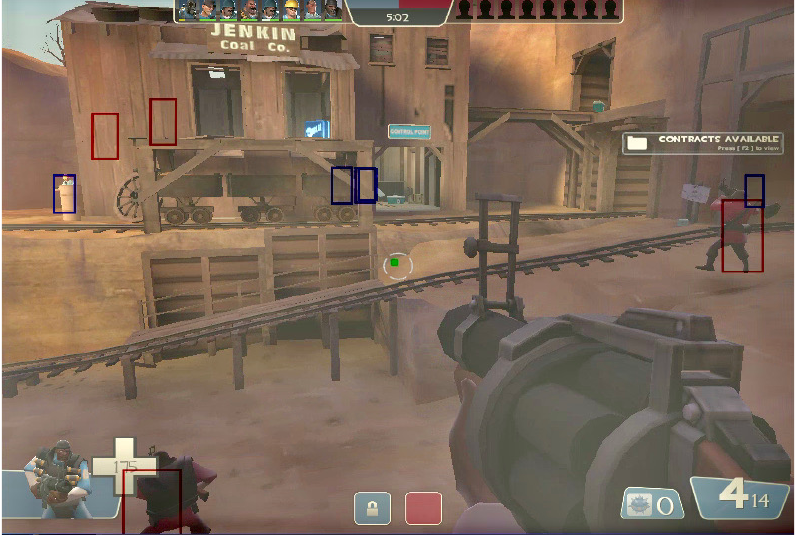}
         \caption{Team Fortress 2}
         \label{fig:tf2-wallhack}
     \end{subfigure}
     \hfill
     \begin{subfigure}{0.45\textwidth}
         \includegraphics[width=\textwidth]{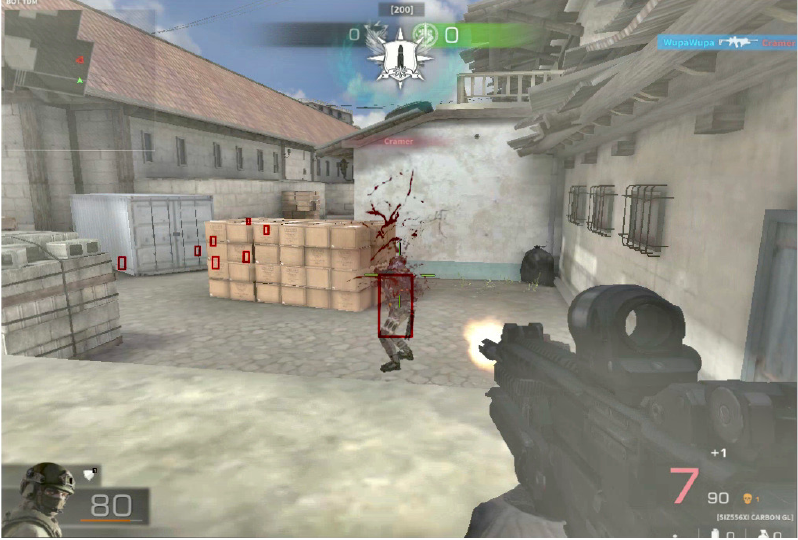}
         \caption{BlackSquad}
         \label{fig:bs-wallhack}
     \end{subfigure}
     \hfill
     \begin{subfigure}{0.45\textwidth}
         \centering
         \includegraphics[width=\textwidth]{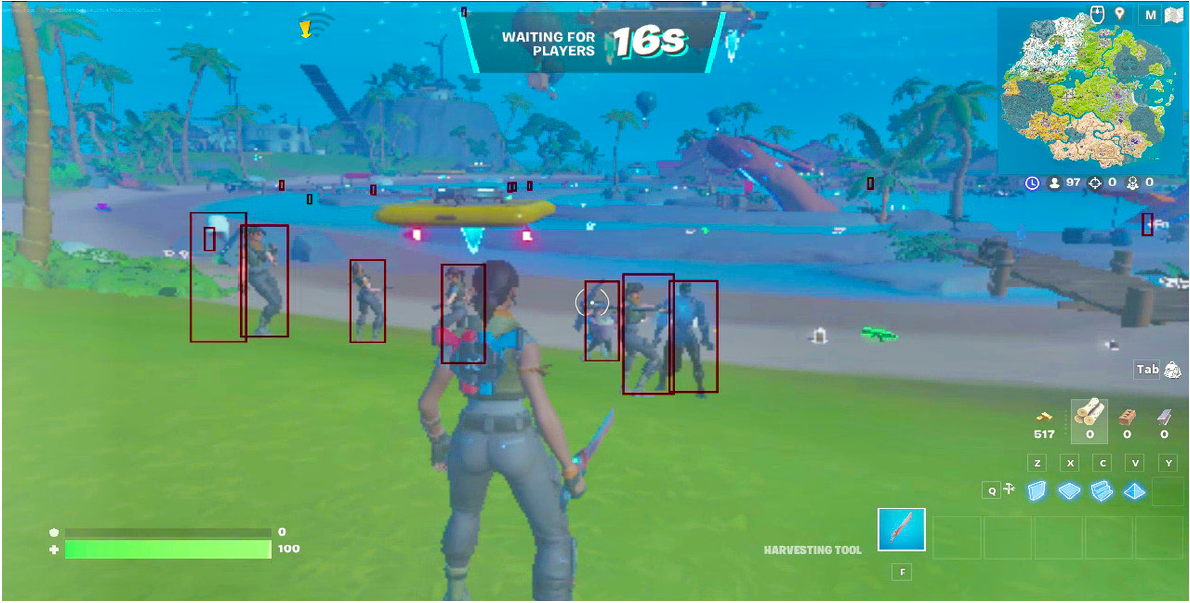}
         \caption{Fortnite}
         \label{fig:frt-wallhack}
     \end{subfigure}
     
        \caption{Wall-hack cheats for each of the three evaluated games using \frameworkName. The wall-hack highlights relevant entities by drawing rectangles around them. In this case, red rectangles represents the enemy avatars while blue ones represent the allies.}
        \label{fig:wall-hacks}
\end{figure}

The Wall-hack starts from the same coordinates obtained for the cheat radar but requires translating the obtained 3D world coordinates to a 2D vector of screen dimensions (as shown on Figure \ref{fig:wall-hacks}). This transformation is referred to as world-to-screen transformation and it is a common design practice in games and game engines. To apply this transformation we need a View Matrix. A View matrix is a 4 x 4 matrix, which transforms world-space vertices to view-space vertices. This transformation is depicted in Figure \ref{fig:world-to-screen}. An example of a transformation matrix can be seen in Equation \ref{eq:tr-matrix}. The first three columns of the matrix represent the orientation of the x,y and z axis in the new space. The fourth column represents the position of the object in the new space. After a view-space transformation has taken place, the objects in view-space are seen from the point of view of the camera lens. To finalize our wall-hack we need to project the view-space coordinates to clip-space using a projection transformation. The projection matrix squeezes the view-space to a two dimensional space, our screen. A more detailed explanation of the world-to-screen transformation can be found in \cite{modelViewProjection}. 

\begin{equation}
\begin{bmatrix}
Or\_X-Axis.x & Or\_Y-Axis.x & Or\_Z-Axis.x & T.x\\
Or\_X-Axis.y & Or\_Y-Axis.y & Or\_Z-Axis.y & T.y\\
Or\_X-Axis.z & Or\_Y-Axis.z & Or\_Z-Axis.z & T.z\\
0 & 0 & 0 & 1 
\end{bmatrix}
\label{eq:tr-matrix}
\end{equation}

\subsubsection{Trigger-bot.\label{subsec:triggetbot}}
Trigger-bots normally execute two consecutive tasks: (i) identify when the local player is aiming at an enemy avatar and (ii) communicate with the local player when to start or stop shooting. Understanding if an enemy is under cross-hair is game design specific, thus it can be achieved by either finding a memory address which stores a ray tracing result of the player's cross-hair or by comparing the screen position of the enemy avatar with the cross-hair position which is usually the screen center. If there is an overlap between these two regions then the enemy is on target. The on target check can vary based on the cheat implementation. We provide two ways of implementing this check behaviour a) with periodic memory reads and b) using a page guard exception which calls a callback that triggers an event e.g. start or stop shooting.

To force the player to star or stop shooting we can either scan the game's process for an address which controls the shooting state and modify (write) the address accordingly when we want to shoot.

A more generic and stealthier solution is to mimic input devices with the help of the VM and use remote interfaces like QMP (QEMU Machine Protocol) messages to send mouse input or keyboard input \cite{qemuQMP}. Mimicking input devices triggers the right methods a normal game input event would trigger. On the other hand changing random bits in memory might trigger some anti-cheat detection measures if that particular address happens, for an example, to be stored in multiple locations and the cheat updated only one of the expected values thus making input device mimicking a stealthier and more abstract method to trigger input events.

\section{Evaluation}
\label{sec:evaluation}

This section describes how we have evaluated our framework against three popular online video games. We first describe the criteria used to select the three games.  We then analyse the performance of the games when the cheats are activated and when they are deactivated. Finally, we discuss the impact, if any, our cheats have on the game's performance and show how they have remained undetected by the anti-cheat systems of the evaluated games.

The three games used for evaluating the VIC framework are TF2, BlackSquad and Fortnite. The games were selected based on the following criteria, summarised on table \ref{table:summaryGameSelectionCriteria}:  
\begin{itemize}
    \item They are first or third person free-to-play action games with an active player base. TF2 and BlackSquad have an average of 95,346 active players and 726 respectively, in the last three months at the time of writing this thesis\footnote{Data taken from \url{https://steamdb.info}, a popular measurement site for the Steam gaming platform}. Fortnite reports 253,998,243 average active players based on data from activeplayer.io \cite{fortniteStats}.
    \item They employ one or more anti-cheat systems. TF2 uses \emph{VAC} \cite{VAC}, Fortnite uses \emph{Byfron} \cite{byfron}, \emph{Easy Anti-Cheat} \cite{EAC} and \emph{BattleEye} \cite{battleEye} and BlackSquad uses BattleEye. VAC stands for Valve anti-cheat and it is a user level anti-cheat, first introduced in 2002 as part of the Steam platform for Counter Strike. Easy Anti-Cheat is a kernel level anti-cheat, it first started as a third party anti-cheat although it was later on acquired by Epic games in 2018 and currently protects a large number of games. BattleEye is a third party kernel level anti-cheat initially released in 2004. Byfron is a user level anti-tamper released in 2022 which makes it the newest of the four. It does not only try to detect hooks on game functions but it introduces obfuscation on the game binary and memory protection during gameplay.
    
    \item They are designed using different game engines, Source Engine, Unreal Engine 3 and Unreal Engine 5 respectively. Game engines provide important building blocks for creating a game, by testing the cheat methodology using different game engines, a better understanding of the performance and efficiency of the proposed attack can be achieved.
    \item They come from different eras but are still being played and frequently updated (at the time of writing this thesis). TF2 was first published on 10th October 2007 (latest update September 2022). BlackSquad and Fortnite represent more recent games as they were published 28th and 21st of July 2017 respectively (latest updates September 2022).
    \item The VM setup used could meet their hardware requirements for smooth gameplay. Smooth play for each game can have a different frame rate as games are designed in different ways based on the engine they use, their genre and finally graphics and network optimisations put in place. One example of genre design difference which impacts performance is the need to support fewer players than other games, TF2 supports up to 24 players but Fortnite supports 100 because it is a battle royal game. What is considered smooth play is a frame rate over 15 fps.
\end{itemize} 

\begin{table}[h]
\centering
\begin{tabular}{lllr}
\toprule
& Team Fortress 2 &  BlackSquad & Fortnite \\
\toprule
Genre &FPS & FPS &TPS \\
Average Player base &95,346 & 726  & 53,998,243 \\
User-level Anti-cheats    & VAC  &N/A &Byfron \\
Kernel-level Anti-cheats & N/A  & BattleEye & BattleEye \& EAC \\
Game Engine &Source Engine  &  Unreal Engine 3 & Unreal Engine 5 \\
Number of players in session  &24 &12 &100 \\           
Release year     &2007 & 2017 & 2017 \\    
\bottomrule
\end{tabular}
\caption{Summary of game selection criteria.}
\label{table:summaryGameSelectionCriteria}
\end{table}

\subsection{Setup specifications}\label{specs}

The experimenters were executed on an Intel i7 7th generation CPU 2.9GHz with 2 cores 2 threads accompanied by Intel's HD Iris 620 GPU and 16GB RAM (8GB allocated to the host and 8GB allocated to the guest). A standard optical USB mouse was used as mouse for the guest machine. The system was running Windows 10 Enterprise OS build 15063, QEMU 2.11.932.11.93 patched for introspection support and KVM Hypervisor version. Team Fortress 2 along with BlackSquad and Fortnite were running the latest version as of 27th September 2022 during the performance testing.

\subsection{Attack Validation} 
\label{attack-validation}
 
 Anti-cheats do not always react upon a cheat detection, they ban in waves. This technique helps the game developers to ban clients of popular cheat makers, as it gives time for the cheats to get shared via communities and create a big negative impact on the cheat developer's reputation \cite{ricochetWaves}. All the cheats designed for the purposes of this paper were tested on private servers when there was an option with the anti-cheat settings activated. In the case where private servers were not an option, the cheats were tested on public matches. Because of the unethical impact the cheats could have on other players' game sessions, the game was abandoned before our plays could affect the outcome of the game. For the purposes of testing the trigger-bots, they were mainly tested in private matches and the cross-hair target was off center when the match was public in order to miss the enemy avatars and avoid detection via server-side behavioural analysis (see \S\ref{sec:discussion} for a more detailed discussion on how to mitigate against this cheat detection techniques). At the time of writing, the cheats had at least an hour of play time each and a number of weeks have passed since testing the cheats live. After combining the ban status of all the accounts used during testing, with the feedback from the vendors through the responsible disclosure process \S\ref{subsec:ResponsibleDisclosure}, non of them have been banned which verifies the stealthiness of our methodology.

\subsection{Performance evaluation}
\begin{figure*}
    \centering
    \includegraphics[scale=0.4]{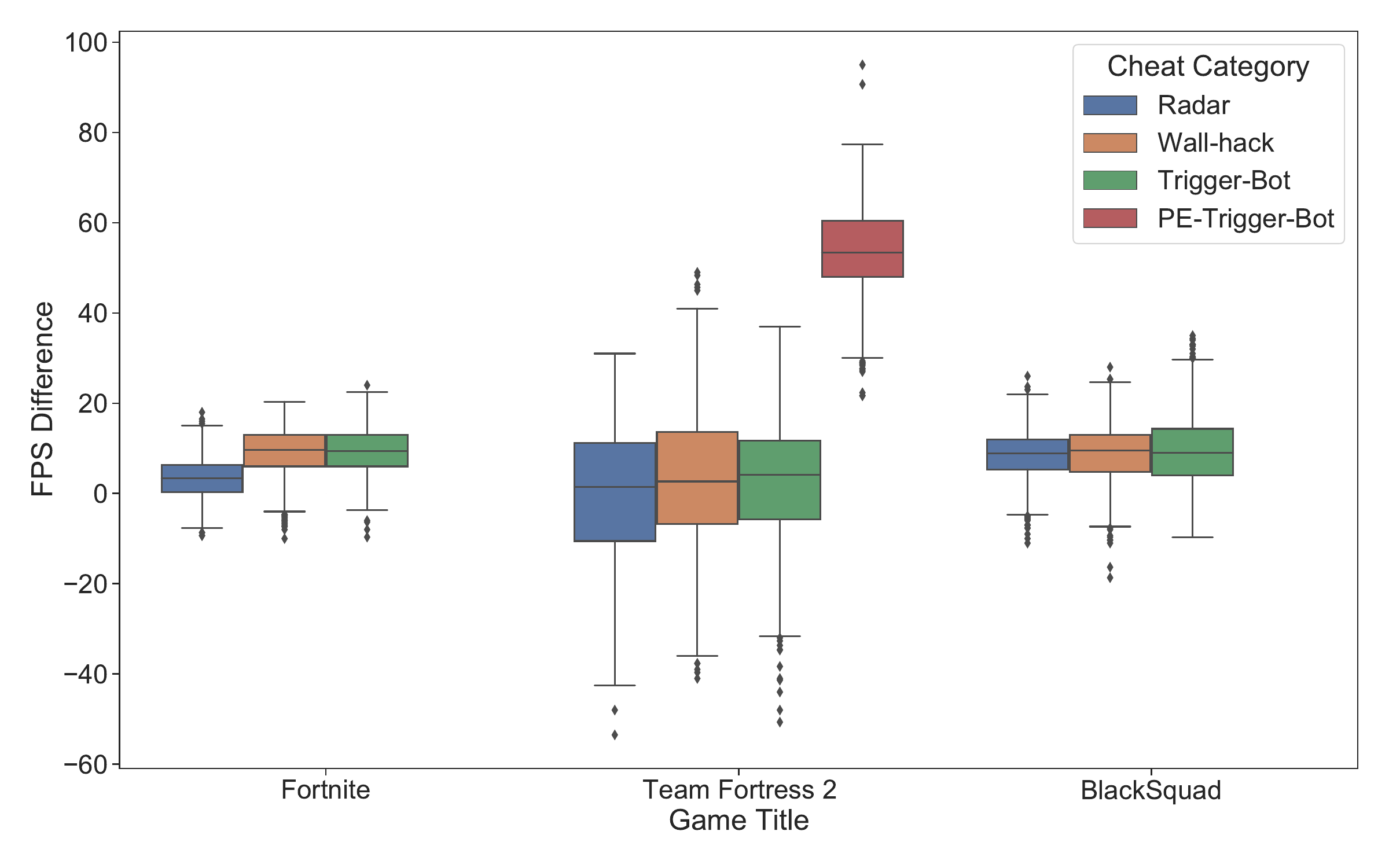}
    \caption{Mean value of FPS difference and quartiles of game sessions under different cheat scenarios.}
    \label{fig:games-fps}
\end{figure*}

Frames per second (FPS) are used very often by PC players as a way to measure the performance of their hardware against the game and vice versa. It is considered the de facto way to test how smoothly the game runs on specific hardware. The higher the frame rate the smoother the scene transitions, thus better reaction times and overall better game experience. Having better reaction times also gives a competitive advantage over other players. This highlights the importance of FPS for both professional and hobbyist gamers. Figure \ref{fig:games-fps} demonstrate the impact our cheats have on the games' performance.

The FPS measurements were collected using FRAPS \cite{fraps} a Windows program for counting FPS, supporting Direct X and OpenGL, used also in the past in other works \cite{fraps::xueffects,fraps::abeysundaraperformance}. We carried out our analysis by monitoring the FPS count of three whole game sessions for each game repeated for every cheat scenario. The duration of a game session was selected by using the default game session time of each game, 600 seconds for Team Fortress 2 and 900 seconds for BlackSquad. In the case of Fortnite, 600 seconds was used as it was the best trade off between using the cheats for enough time to prove they are stable and being able to stay alive without killing people thus tampering with the session's outcome.

We start by measuring the FPS count without running any cheats to establish a benchmark. Then, the same steps were repeated for every scenario a) cheat-radar, b) wall-hack, c) trigger-bot with periodic reads and d) trigger-bot using memory events (only for TF2\footnote{The authors could not identify a similar address in the other two games responsible for reflecting the object under crosshair. This can occur because of different game design decisions.}). Each cheat scenario was monitored three times. From those three sessions we constructed an average scenario session by averaging the FPS count for each second. As a final step the average benchmark session was subtracted from the cheating sessions. The mean and quartiles of the difference of every cheat experiment are depicted in Figure \ref{fig:games-fps}. The y axis reflects the FPS difference with positive numbers representing how many frames the cheat was slower than the normal game operation and negative numbers how much faster. 

The performance of TF2 was measured at 60fps without any cheats activated. Cheat radar had an identical average performance with the benchmark session at 60fps followed by Wall-hack and Trigger-Bot with an average performance of 57fps. The lowest performance was reported by the page guard trigger-bot at 5fps.

The event implementation of the trigger-bot creates a performance reduction as it limits the FPS average output to 5 fps. This happens because the monitoring page frame triggers a few hundreds of events per second irrelevant to the address we are monitoring. There exist a correlation, on the examined game, between the number of memory events and the total number of players in the game session. When the number of players gets increased the number of memory events also increases. This creates a performance impact on the VM as it needs to process these events, even-though they are irrelevant. A way to overcome this limitation is by using a Sub-page write protection (SPP) event, supported by Intel 10th generation or later CPU's as described in \S28.3.4.2 of the Intel Developer's Manual \cite{IntelSPP}. This provides the ability to the user to monitor a sub range of addresses in a page frame thus eliminating the interception on page execute events irrelevant to the cheat. The use of different engines and the implementation differences did not provide us the opportunity to test the PE-Trigger-Bot on BlackSquad and Fortnite as the authors could not find a memory address reflecting the aim state of the local avatar.

Figure \ref{fig:games-fps} shows the performance comparison of VIC on BlackSquad. The game runs with an average frame rate of 60fps on our system. It is important to mention because of the use of the integrated graphics the game was set up to run on low graphics performance setting with a maximum frame rate of 60fps. The impact of the cheats on the game's performance is negligible. Our results report the lowest average frame rate to occur while running the Wall-hack cheat at 50fps only 10 frames below the average frame rate performance when having no cheats enabled. A number of outliers can be spotted across all cheat categories the lowest being at 25fps while running the trigger-bot. The reason for this is due to the graphics card's power. Passing a GPU to the guest VM will diminish these slight drops in performance and increase the average frame rate of the game. We did not test a GPU pass through setup because of hardware limitations of the equipment used for the experiments (e.g. the experiments were carried on a laptop). All cheats apart from trigger-bot with page guard events have negligible performance impact on the games tested.

The third and final game we tested was Fortnite, because of the game's nature and requirements it was clocked at an average of 25fps on our system. Despite the relatively low fps, because of in game network optimizations it was running smoothly. Figure \ref{fig:games-fps} shows the performance comparison of our cheats on Fortnite. In contrast to Team Fortress 2, Fortnite and BlackSquad both employee Unreal Engine 5 and Unreal Engine 3 respectively. 

Our experiments show that the cheat radar introduced via \frameworkName had a negligible effect in the three games we tested (average of 3fps decrease). The wall-hacks and trigger-bots have an average performance loss of 10 and 8 fps respectively. We attribute this to two main facts. First, these games run very similar game engines, which probably means that external factors like a higher CPU load may also affect the game engines on a similar way. Second, both the wall-hack and trigger-bots were designed on a similar way, requiring more complex operations to calculate the overlay and the time to pull the trigger respectively compared to the cheat radar implementation.

\section{Mitigations \& Discussion}
\label{sec:discussion}

While we were able to execute cheats via \frameworkName without being detected, the fact that they execute via a hypervisor could be used to help in their detection. This section discusses the different mitigations proposed in the literature in this direction. We also discuss how these new kinds of cheats could affect the current business model of cheat developers. In particular, we talk about the potential impact of cloud gaming in regards to cheating in online video games.

\subsection{Hypervisor detection}

\subsubsection{\textbf{Timing techniques}}
From the point of view of the game developer and the integrity of the game, cheating can be considered a malicious act. Thus, the hypervisor can be characterised as a malicious introspection hypervisor. Lots of research was conducted on detecting the use of a hypervisor although as highlighted by Tuzel et al. in \cite{whoWatchesTheWatcher} the important distinction of detecting an introspected hypervisor from a hypervisor becomes more and more important and relevant as time progresses. In their paper, they introduce ways to detect introspection using different indicators. \textit{Instruction intercession}, \textit{Active memory intercession} and \textit{Passive memory monitoring}. 

Daax et al. have researched how anti-cheats try to deal with the problem of cheating using a hypervisor. To be more specific, they reverse engineered BattleEye \& Easy Anti-Cheat and give a great inside on how these anti-cheats detect emulation \cite{DaaxAntiCheatEmulation}. They report that both anti-cheats use a standard timing attack using the \textbf{\textit{rdtsc; cpuid; rdtsc}} instruction combination to count the duration it takes to execute the cpuid instruction as it triggers a vmexit. This can be avoided by using a proper TSC emulation at the hypervisor level. Easy Anti-Cheat was reported to perform an additional check. 
 It executes the  \textit{\textbf{vmread}} instruction upon driver initialisation and checks for a \textit{\#UD} (undefined instruction exception).
 
 They conclude these checks are simple to circumvent (at the time of writing their report in 2020). The experiments carried out for this work validate their findings. No extra steps were taken to hide the hypervisor during the experiments. Tuzel et al. \cite{whoWatchesTheWatcher} also comment on the unreliability of the TSC clock and recommend the use of the x86 HPET timer as an alternative. HPET is less precise compared to TSC, although it is more difficult to manipulate using a hypervisor, as it can introduce problems to the guest's behaviour. It can cause problems with video and audio playback, which is undesirable when playing a video game \cite{intelHPET}.

 \subsubsection{\textbf{Page Guard Exceptions}}
The VIC framework utilises page guard exceptions to establish a monitoring mechanism capable of observing changes in address values, as detailed in \S \ref{sec:building_blocks}. The monitoring mechanism is a cheat implementation improvement as the cheat is notified by an exception callback handler on the hypervisor level when a monitoring address changes. A way to defend against this mechanism is to increase the number of callbacks that the hypervisor needs to process. This will harm the cheat performance, thus removing any advantage the cheater will gain over other players while using it.  

In response to this mechanism, a game developer can leverage the use of huge pages, a memory page of 2MB or 1GB in size instead of 4KB, to allocate memory in the game process. A bigger page consists of a larger number of addresses; thus, allocating important addresses to this type of page will introduce a large number of interrupts and will cause a VIC cheat with the monitoring callback behaviour to suffer performance degradation. Huge page support depends on the OS and hardware support.

\begin{figure}
\centering

\begin{lstlisting}[language=C, basicstyle=\small\ttfamily]
SIZE_T pageSize = 4096; // 4KB page frame
LPVOID pageMemory = VirtualAlloc(NULL, pageSize,
                              MEM_COMMIT, PAGE_READWRITE);

// Example: NOP instruction (0x90) for x86 and x86-64
unsigned char codeInstruction[] = {0x90}; // NOP
int gameVariable = 5;

// Copy the code instruction into the allocated memory
memcpy(pageMemory, codeInstruction, sizeof(codeInstruction));

// Copy the data variable into the allocated memory, 
// right after the code instruction
memcpy((void*)((SIZE_T)pageMemory + sizeof(codeInstruction)),
                        &gameVariable, sizeof(gameVariable));

__try {
    ((void(*)())pageMemory)();
} __except (GetExceptionCode() == EXCEPTION_ACCESS_VIOLATION ? 
                                  EXCEPTION_EXECUTE_HANDLER :
                                  EXCEPTION_CONTINUE_SEARCH) 
{
    // Change permission to execute
    VirtualProtect(pageMemory, pageSize, 
                   PAGE_EXECUTE_READ, 
                   &oldProtect);
    
    // Re-execute the instruction now that it's allowed
    ((void(*)())pageMemory)(); 

    // Revert permissions to induce faults again
    VirtualProtect(pageMemory, pageSize,
                   PAGE_NOACCESS, &oldProtect);
}
\end{lstlisting}
   
\caption{Pseudo-code of Windows API code allocating a page memory consisting of code and data.}
\label{fig:pageFault-code}
\end{figure}

A cheater may choose to disable this feature at the hypervisor or OS level, to safeguard against the utilisation of huge pages. This action, however, prevents the system from accessing the numerous benefits that come with the use of huge pages. For instance, one of the key advantages is a reduction in cache misses, a direct result of the translation lookaside buffer (TLB) holding fewer entries. The TLB, a critical memory cache, is responsible for storing the recent translations from virtual to physical memory. Additionally, another significant benefit is the ability of the system to avoid swapping pages outside of the main memory, thereby maintaining operational efficiency and improving performance \cite{benefits-huge-pages}.

An increase in the total number of page faults can also be achieved by the allocation of both code and data within the same page frame, resulting in a fault being generated with each execution of the code instruction allocated. This scenario is illustrated in Figure \ref{fig:pageFault-code}, where a memory page is allocated and populated with a \textit{NOP} instruction. Following this instruction in memory, an integer is placed, representing a game variable that the developer aims to safeguard. The latter part of the code illustrates the execution of the instruction from within the game process, followed by a try-catch statement. It is demonstrated by the catch statement's exception handler how the permissions of the memory page are altered to execute the instruction, before being reverted to \textit{PAGE\_NOACCESS} permission, thereby reinstating the page guard exception for future occurrences.

Both of the above techniques aim at increasing the number of exceptions/interrupts passed to the VIC callback system. As already discussed in \S \ref{attack-validation} the use of SPP available on some Intel CPUs can alleviate a big part of the extra exceptions created by the use of huge page frames and the construction of page frames which include neighbouring code instructions and game data. SPP increase the granularity of monitoring a sub-page to 128 bytes as described in Intel's Software Developer's manual \cite{IntelSPP}.
\subsection{Process Enclaves}
In recent years, the academic community have also proposed anti-cheat solutions which employ the use of enclaves like Intel SGX to give an answer to the on going issue of cheating  \cite{sgxCheatDetection,sgxBlackmirror}. These solutions seem promising because of their anti-tamper properties, although even without taking account the number of emerged attacks on enclaves; there are a few more issues to consider. Intel discontinued SGX for their consumer chips at the time of writing this paper \cite{intelSGXComm}. This introduces a discrepancy as game developers cannot force their audience to use specific hardware in order to be able to play a game. Another important factor is the extra development time added by these custom hardware solutions, as they increase the support and development time for the game security teams or anti-cheat developers. Mainly because they have to support different manufactures on top of different OS support if the game happens to be available across different platforms e.g. Linux, Windows and MacOS. In this regard, Google Cloud introduced Asylo in 2018, an enclave development framework which aims to help tackle the problem of supporting multiple enclave applications \cite{googleAsylo}.

\subsection{Domain Enclaves}

In contrast to process based enclaves (Intel-SGX) that try to protect parts of a process by running it in an enclave; Intel-TDX and AMD-SEV offer a domain based confidential environment by placing the whole virtual machine in an enclave \cite{IntelTDX}. This creates isolation between other protected VMs and a malicious hypervisor. The VM memory is encrypted and the hypervisor has only access to necessary information needed to keep the VM running. This is achieved by placing a trusted hardware module between the hypervisor and the VM which takes the responsibility to handle interrupts and memory events. Domain enclaves have a great advantage over process enclaves as they do not require any application modification to be used.

The thread model of both Intel-TDX and AMD-SEV fits perfectly with the cheating framework we propose in this paper and can be a good solution to the problem. Unfortunately at the time of writing Intel-TDX is not available on any hardware yet and AMD-SEV is available only with AMD EPYC CPUs which are specifically targeting server and embedded system markets, not consumer devices or gaming PCs \cite{AMD-SEV}.

\subsection{Community impact}
\label{sec:discusion:business_model}

Game cheating communities are organised in online forums and private online groups, as discussed in Karkallis et al. \cite{DetectingGameInjectors}. At the time of writing, UnkownCheats and MPGH have 5 million and more than 4 million members, respectively. VIC cheats impact the way cheats are exchanged in these communities; as a pre-setup of the hypervisor and introspection libraries is required to be in place before running the cheats. The current exchange model usually involves getting access to cheat forums or private discord groups, paying or downloading free cheats in the form of an executable or DLL and injecting them into game memory in the case of DLLs.

The proposed cheats are more difficult to set up, as the cheat developer will have to provide additional preinstalled scripts or guides for the installation and setup of the hypervisor and introspection libraries. The rest of the process will remain identical, as the cheat developer can still compile the actual cheat as an executable for download, although the cheat's interactions with memory and interrupt signals need to be manually modified, as they need to use the hypervisor and run on the host machine as explained in \S \ref{sec:vmiCheat}. In the future, cheat developers might automate this transition by abstracting the cheat interactions. They can also provide a VM for download with the entire setup. This could also eliminate the cheating issues that arise from different OS versions. 

The use of VMs is also strongly coupled with new trends like cloud gaming. Many cloud providers have taken a step towards cloud gaming, with slightly different business models. Some examples are Nvidia's GForceNow and Xbox Cloud Gaming. Cloud gaming offers the advantage of running the actual game remotely and reduces the attack vector of cheats to screen and input devices. During the writing of this thesis, one of the major cloud players, Google, discontinued their cloud gaming service, Stadia \cite{stadiaClosure}. This might be an indicator that cloud gaming technology is immature at the time of writing this thesis. On the other hand, this might not be enough to stop cheat developers from creating Game-Cheating as a service (GCaS) and offer cheaters remote access to VMs with cheats enabled. 

This way cheaters would not need any cheating expertise to access a cheat. They will just connect to the GCaS guest which will be pre-configured with a VMI-enabled cheat and just play the game. This would also reduce some of the risks associated with game cheating, as they will no longer run the risk of installing a potentially malicious program embedded within a cheat. A great example of this is the Remote Access Trojan (RAT) disguised as a Call of Duty cheat in 2021 \cite{callOfDutyTrojan}. In addition, using CGaS allows cheaters to avoid hardware bans (which identify and ban a particular device ID). The game would be running inside the VM on the GCaS service, transferring this responsibility to the cheat developer and platform maintainer. The cheaters would still have to create a new account if their account is blocked.

\subsection{Responsible disclosure}
\label{subsec:ResponsibleDisclosure}
The authors are responsible for disclosing and informing any parties affected by the outcome of this research. The author contacted the game developers of Fortnite, TF2 and BlackSquad on 15th December 2022. The author informed the game developers about the possibility of bypassing their current anti-cheats via VMI. The chapter contents were attached to the initial submission which was carried via direct communication with the game developer or through their bug bounty programs when one was available. There was a follow-up correspondence on 4th January from the Fortnite team. They asked to play at specific times with the cheats activated to gather more data. They later applied a fix by not letting the game start in a VM. When asked about an official update on the issue later on there was no reply back from them after numerous attempts. 

\subsection{Limitations}
\label{sec:limitations}

The \frameworkName framework shares similar limitations with traditional cheats. It is bound by behavioural analysis carried out on the game server by employing ML models \cite{pinto2021deep,tao2020xai,spijkerman2020cheat} or player complaint reports \cite{eaCheatReporting}. This means a cheater can only be as good as a professional player if they do not want to be detected by statistical detection vectors. 

The cheat implementations discussed in this chapter do not have direct access to in game functions. An advantage internal cheats have over the proposed cheat methodology and external cheats, in general. This limitation can be mitigated by copying and emulating the assembly code consisting of an in game function on a vmexit on the host. 

The use of page guard exceptions offers an elegant way of monitoring memory changes. However, as demonstrated in \S \ref{sec:evaluation} they suffer a big performance hit if the address being monitored happens to co-exist on an event busy page frame.  This can be mitigated with EPT-Based SPP, as previously discussed, when the hardware supports it. Another important thing to consider is the need to identify an address that represents a state useful to the cheat. This is game design-dependent. 

The assumption that video games are security-agnostic is made in the threat model \S\ref{sec:threatModel}. Thus the use of simple hacks like infinite health or one shot kill hacks is not possible. A securely designed game should validate these states on the server side and not on the client side which will make the use of these types of cheats impossible.

There also exist some feature oriented limitations. VMI is still maturing and supports specific hardware, mainly Intel CPUs, and is not widely available for all OSs at the moment of writing.

The methodology uses a hypervisor to hide the cheat's actions. In the scenario where the memory of the guest is encrypted, the host will not be able to make sense of what is returned from a physical memory read. Thus, encrypting the game memory can be problematic for the methodology. This work is limited only to hardware which supports virtualisation as the use of a hypervisor is a mandatory part of the methodology.

\section{Conclusions}
\label{sec:conclusions}

In this paper, we presented \frameworkName a new type of online video game cheating methodology which employees virtualization and introspection to bypass modern anti-cheats. We tested the effectiveness of the attack on three different games including Fortnite, each of them running in between them five different anti-cheat systems. We evaluated the performance of our cheats for every game and reported negligible impact on the games' performance. 

Cheating in online video games remains an open problem despite the already proposed solutions. Our methodology can provide cheaters an unfair advantage over other players while remaining undetected. While this kind of cheat is more complex to setup, it could enable cheat developers to build cheat-enabled cloud gaming services (CGaS). We hope that this paper can raise awareness among game and anti-cheat developers and foster the development of new techniques to identify these kinds of cheats, as we believe they will become more prevalent as the usage of cloud gaming becomes more common among players.








\end{document}